\documentclass[preprint,3p,authoryear]{elsarticle}
\usepackage{graphicx,dcolumn,bm,amssymb,amsmath,ulem,subfig,indentfirst,amsthm,caption}
\usepackage{color}

\theoremstyle{plain}
\def\be{\begin{equation}}
\def\ee{\end{equation}}

\newtheorem*{theorem*}{Theorem}

\begin{document}
\author{Bingyu Cui}
\author{Eugene M. Terentjev}
\ead{emt1000@cam.ac.uk}

\address{Cavendish Laboratory, University of Cambridge, JJ Thomson
Avenue, CB3 0HE Cambridge,
U.K.}

\begin{abstract}
We review and compare the Born-Huang and the Lemaitre-Maloney theories that lead to analytical expressions for elastic constants, accounting for affine and nonaffine deformations in a lattice (or in a disordered solid). The Born-Huang method is based on Helmholtz free energy while the Lemaitre-Maloney formalism focus on the Gibbs ensemble with the focus on local force. Although starting from different perspectives, in the linear elastic limit, and in equilibrium, material elastic constants must be the same in all these methods. This is explicitly verified on examples of linear chains, and the numerical simulation of a non-centrosymmetric crystal.
\end{abstract}


\title{Comparison of the Helmholtz, Gibbs, and collective-modes methods to obtain nonaffine elastic constants}
\maketitle

\section{Introduction}
It is manifest that elastic materials experience internal resistance to the deformation caused by external forces. They tend to return to original sizes and shapes when the external influence is eliminated. The elasticity of materials is generally described by a stress-strain curve, which exhibits a characteristic linear region for sufficiently small deformations. This linear regime is vital for, e.g. elastic waves, and most elastic theories are established in this linear regime. In a one-dimensional rod, the simplest linear relation between stress and strain is known as Hooke's law; in three dimensions, the general proportionality between stress and strain is a 4th-rank tensor of stiffness coefficient~\citep{Landau3}. \\

At zero temperature, once the relative initial positions of atoms are known, it is then a simple task to add all contributing interactions to elastic constants for homogeneous (affine) deformations. The resultant elastic constants are often called affine. When the two assumptions,  zero temperature and homogeneous displacements, are not valid, one needs to develop a more complicated theory of nonaffinity (local, inhomogeneous). Early works focus on thermal effects on elasticity in crystals~\citep{Squirre1968,Hoover1969}. In recent decades, athermal systems, like granular materials or foams, raise a lot of attention, investigating corrections to the affine elasticity~\citep{Lacasse1996,Langer1997,Radjai2002,Wittmer2002,Tanguy2002}. In other words, even at zero temperature, particles (atoms) do not always follow homogeneous displacement fields. They instead attempt to minimise the potential energy of the system, and in some cases, this requires additional local nonaffine displacements, no matter how small deformation the system is strained to. The nonaffine correction to the elastic constants can be prominent, which has been found in simulation of a non-centrosymmetry lattice~\citep{Cui2019nonaffine}. The formal expressions for the nonaffine corrections were systematically developed by Lemaitre and Maloney (LM) via studying the Gibbs ensemble with the local force acting on each particle in the system~\citep{Lemaitre2006}. Through performing normal mode decomposition, their analysis relates nonaffine corrections to the correlator of a fluctuating force field, which can be extended to the viscoelastic dynamical response of the system.

Prior to LM, the linear elastic constants were studied in detail in the work of Born and Huang (BH). The most familiar BH results are for the basic affine elastic constants, although they have actually discussed the nonffine deformation case in great detail (but have not derived complete analytical expressions for nonaffine corrections)~\citep{Born1952}. However, reviewing the BH theory, and comparing it with LM formalism, we find that they address the elasticity problem from two complementary angles: LM approach works by identifying the local nonaffine forces, essentially working in the Gibbs ensemble framework, while BH arguments are based on optimization of local nonaffine displacements (i.e. in the conjugate Helmholtz ensemble). To test the comparison between these two approaches, we also consider the vibrational lattice waves in the lattice, which represent collective motions: in the long-wavelength limit these waves provide an additional path to material elastic constants, which must match both of LM and BH results.

This paper is organised as follows: Section 2 reviews the three approaches to elastic constants, including their interpretations of both affine and nonaffine contributions. Section 3 begins with clarifying the link between Gibbs and Helmholtz frameworks, with supporting examples of 1D linear lattices and a 3D non-centrosymmetric crystal, where we compare in detail the different ways of calculating elastic constants. Finally, in Section 4, we draw our conclusions and suggest an insight of practical applications of these methods.

\section{Review of different approaches to elastic constants}
\subsection{BH: elastic constants for non-ionic crystals}  \label{sec2-1}
In non-ionic crystals, only short-range pairwise interaction need to be considered. The long-range Coulombic forces that usually cause a notional divergence are ignored here, although there exist ways to tackle the issue of divergence (see Section \ref{sec3-2} below). \textbf{To make it convenient for calculation, we assume the pair interacting potential depends on the square of interparticle distance near its equilibrium. We also require the system we are studying in the paper remains at zero temperature without internal tension, so that the system will stay in mechanical equilibrium.} We take into account linear deformation in a small neighborhood of a material point in the reference frame $\mathbf{R}^0$:
\begin{equation}
s^\mu(\mathbf{R}^0+\delta\mathbf{R})=s^\mu(\mathbf{R}^0)+\sum_\nu\frac{\partial s^\mu}{\partial R^\nu}\delta R^\nu, \label{eq1}
\end{equation}
where $\mu,\nu$ label the Cartesian vector components. The first term in the RHS represents the translation of the body as a whole, while the last term is the elastic deformation. In a lattice, the particle (atom) $I$, which belongs to the cell $l$, has a continuous displacement $s_I^\mu(l)$. Equation \eqref{eq1} is then equivalent to
\begin{equation}
s^\mu_I(l)=s^\mu_I+\sum_\nu\frac{\partial s^\mu}{\partial R^\nu}R^\nu_I(l) \label{eq2}
\end{equation}
where $s_I^\mu$ is the additional (nonaffine) displacement of a particle $I$ inside the cell in response to the external homogeneous elastic deformation. This nonaffine displacement is periodic, that is, the nonaffine re-arrangement is the same in all cells in the body. From this, we have $s^\mu_J(l')-s^\mu_I(0)=s^\mu_J-s^\mu_I+\sum_\nu (\partial s^\mu/\partial R^\nu)R^\nu_{IJ}(l')$. By $R^\nu_{IJ}(l')$, we mean the relative displacement $R^\nu_{J}(l')-R^\nu_I(0)$.
In general, there are $N$ particles in each cell and $d$ is the space dimension. As is shown in the Appendix A, the change in energy (density) per volume of unit cell $v_a$ can be written as
\begin{align}
&U=\frac{1}{2}\sum_{IJ\mu\nu}\left\{\begin{matrix}
I &J\\  \mu &\nu  \end{matrix}\right\}s^\mu_I s^\nu_J
+\sum_{I\mu\nu\xi}\left\{\begin{matrix}  I &\nu\xi\\ \mu \end{matrix}\right\}s^\mu_I \frac{\partial s^\nu}{\partial R^\xi}
+\frac{1}{2}\sum_{\mu\nu\xi\iota}\{\mu\nu\xi\iota\}\frac{\partial s^\mu}{\partial R^\nu}\frac{\partial s^\xi}{\partial R^\iota}, \quad\text{where}\label{eq3U}\\
&\left\{\begin{matrix}
I &J\\
\mu &\nu
\end{matrix}\right\}=\frac{2}{v_a}\left\{\delta_{IJ}\delta_{\mu\nu}\sum_{l'K}[V']_{\mathbf{R}_{IK}(l')}-\delta_{\mu\nu}\sum_{l'}[V']_{\mathbf{R}_{IJ}(l')}+2\delta_{IJ}\sum_{l'K}[V''R^\mu R^\nu]_{\mathbf{R}_{IK}(l')}-2\sum_{l'}[V''R^\mu R^\nu]_{\mathbf{R}_{IJ}(l')}\right\};\notag\\
&\left\{\begin{matrix}
I &\nu\xi\\
\mu
\end{matrix}\right\}=-\frac{4}{v_a}\sum_{l'J}[V''R^\mu R^\nu R^\xi]_{\mathbf{R}_{IJ}(l')};\notag\\
&\{\mu\nu\xi\iota\}=\frac{2}{v_a}\sum_{l'JI}[V''R^\mu R^\nu R^\xi R^\iota]_{\mathbf{R}_{IJ}(l')}. \notag
\end{align}

Denoting the {linear} (external, symmetric) strain tensor $\underline{\underline{\eta}}$ by
\begin{align}
\eta_{\mu\nu}= \frac{1}{2}\left(\frac{\partial s^\mu}{\partial R^\nu}+\frac{\partial s^\nu}{\partial R^\mu}\right), \label{eq4strain}
\end{align}
the energy density $U$ is rewritten as
\begin{align}
U&=\frac{1}{2}\sum_{IJ\mu\nu} \left\{\begin{matrix} I &J\\ \mu &\nu \end{matrix}\right\}s^\mu_I s^\nu_J
+\sum_{I\mu\nu\xi}\left\{\begin{matrix} I &\nu\xi\\ \mu \end{matrix}\right\}s^\mu_I\eta_{\nu\xi}
+\frac{1}{2}\sum_{\mu\nu\xi\iota}\{\mu\nu\xi\iota\}\eta_{\mu\nu}\eta_{\xi\iota}  \label{eq5U2}
\end{align}
Physically, the (internal) nonaffine displacement $s_I^\mu$ adjusts such that the energy density becomes minimum for the given external elastic strain components $\eta_{\mu\nu}$. That is,
\begin{align}
0=\frac{\partial U}{\partial s^\mu_I}=\sum_{J\nu}\left\{\begin{matrix} I &J\\ \mu &\nu \end{matrix}\right\}s^\nu_J
+\sum_{\nu\xi}\left\{\begin{matrix} I &\nu\xi\\ \mu \end{matrix}\right\}\eta_{\nu\xi},
\label{Bornequilibrium}
\end{align}
which gives $(N-1)d$ independent equations. The solutions $s_I^\mu(\underline{\underline{\eta}})$ of these mechanical equilibrium conditions are, in fact, the nonaffine displacements.
Since the energy density depends only on the differences between the different $\mathbf{s}_I$, without loss of generality, we can let $\mathbf{s}_1=0$ after an arbitrary shift of the reference frame origin. When the internal displacements are eliminated as independent variables, the energy density becomes a quadratic expression in $\underline{\underline{\eta}}$, whose coefficient matrix $\{\mu\nu\xi\iota\}$ will receive a correction after substituting the solutions for $\mathbf{s}_I$, which we denote as $\{\mu\nu\xi\iota\}'$.
The detailed analysis of the resultant correction to elastic constants is shown in Section 3.
In the regime of linear elasticity, we define the stress tensor as
\begin{align}
\sigma_{\mu\xi}=\sum_{\nu\iota}C_{\mu\xi\nu\iota}\eta_{\nu\iota}\equiv\sum_{\nu\iota}\{\mu\xi\nu\iota\}'\eta_{\nu\iota}. \label{eq7stress}
\end{align}
This stress tensor $\sigma_{\mu\xi}$ represents the $\mu$-component of the force exerted on the medium which is on the negative of a unit surface segment normal to the $\xi$-direction. Thus, its divergence is the local force per unit volume. The local equation of motion is then expressed as,
\begin{equation}
\rho\ddot{s}^\mu=\sum_{\xi}\frac{\partial\sigma_{\mu\xi}}{\partial R^\xi}=\sum_{\nu\xi\iota}C_{\mu\xi\nu\iota}\frac{\partial^2s^\nu}{\partial R^\xi\partial R^\iota},  \label{eq8wave}
\end{equation}
with mass density $\rho$.
To solve this, we substitute the elastic wave with an amplitude vector $e^\mu(\mathbf{q})$:
\begin{align}
&s^\mu(\mathbf{R},t)=e^\mu(\mathbf{q})e^{i\mathbf{q\cdot R}-i\omega t}\notag\\
\Rightarrow\quad&\rho\omega^2e^\mu(\mathbf{q})=\sum_\nu\left(\sum_{\xi\iota}C_{\mu\xi\nu\iota}q^\xi q^\iota\right)e^\nu(\mathbf{q}). \label{eq9wave2}
\end{align}
This is the equation for elastic waves in solids.

\subsection{Long-range acoustic waves from collective modes} \label{sec2-2}
The review of lattice vibrations  is provided in the Appendix B.
Now we assume the particle $I$ carries mass $m_I$, and denote the local force component $f^\mu_I(l)=-\left(\partial\mathcal{U}/\partial s^\mu_I(l)\right)_{R_0}$, with the corresponding Hessian matrix $H^{\mu\nu}_{IJ}(l l')=\left(\partial^2\mathcal{U}/\partial s^\mu_I(l)\partial s^\nu_{J}(l')\right)_{R_0}$. Here, $\mathcal{U}$ is the total potential energy, directly related to the energy density used before, $\mathcal{U}=U\cdot V$ where $V$ is the volume of the whole system. Physically, $f^\mu_I(l)$ is the total force on the atom $(l,I)$ in the reference configuration $R_0$, while $-H^{\mu\nu}_{IJ}(l l')$ is the linear response coefficient in the $\mu$-component of force acting on the atom $(l,I)$ due to the displacement of an atom $(l',J)$ along $\nu$-direction.

As shown in Appendix B, by expanding the potential energy with respect to the displacement of each particle up to the 2nd order, when the position of particles changes from $R_I^\mu(l)$ to $R_I^\mu(l)+s_I^\mu(l)$, the equation of motion becomes:
\begin{equation}
m_I\ddot{s}^\mu_I(l)=-\sum_{l'J\nu}H^{\mu\nu}_{IJ}(l l')s^\nu_J(l').
\label{latticeeom}
\end{equation}
We look for wave solutions of this equation, with $s^\mu_I(l)=e^\mu_Ie^{i\mathbf{q\cdot R}_I(l)-i\omega t}/\sqrt{m_I}$, where $\mathbf{q}$ is an arbitrary vector in the reciprocal space ($|q|$ is the wave number).
Substituting this solution into Eq. (\ref{latticeeom}) gives
\begin{align}
&\omega^2(\mathbf{q},j)e^\mu_I(\mathbf{q},j)=\sum_{J\nu}\mathcal{M}^{\mu\nu}_{IJ}(\mathbf{q})e^\nu_J(\mathbf{q},j), j=1,2,...,Nd.
\label{eomexpandq}
\end{align}
with the dynamical matrix defined as
\begin{align}
\mathcal{M}^{\mu\nu}_{IJ}(\mathbf{q})=\frac{1}{(m_Im_J)^{1/2}}\sum_{l'}H^{\mu\nu}_{IJ}e^{-i\mathbf{q\cdot R}_{IJ}(l-l')}.
\end{align}
Thus, for each $\mathbf{q}$, there exists $Nd$ solutions of $\mathbf{s}_I(l)$.
It is easy to verify the symmetry properties of the dynamical matrix:
\begin{equation}
\mathcal{M}^{\mu\nu}_{IJ}(\mathbf{q})^*=\mathcal{M}^{\nu\mu}_{JI}(\mathbf{q})=\mathcal{M}^{\mu\nu}_{IJ}(-\mathbf{q}).
\end{equation}

Looking for the effective elastic constant, let us consider a small perturbation in the wavevector from $\mathbf{q}=\mathbf{0}$ along one of the 1D acoustic branches. Introducing the small scalar number $\epsilon$ to track the orders of magnitude of the small vector $\mathbf{q}$, we have:
\begin{align}
\mathcal{M}^{\mu\nu}_{IJ}(\epsilon\mathbf{q})&=[\mathcal{M}^{\mu\nu}_{IJ}]^{(0)}+i\epsilon\sum_{\xi}[\mathcal{M}^{\mu\nu,\xi}_{IJ}]^{(1)}q^{\xi}+\frac{\epsilon^2}{2}\sum_{\xi\iota}[\mathcal{M}^{\mu\nu,\xi\iota}_{IJ}]^{(2)}q^{\xi}q^{\iota}+...\label{perDA}\\
\label{perwA}
\omega(\epsilon\mathbf{q},j)&=\epsilon\omega^{(1)}(\mathbf{0},j)+\frac{\epsilon^2}{2}\omega^{(2)}(\mathbf{0},j)+...\\
e^\mu_I(\epsilon\mathbf{q},j)&=[e^\mu_I(\mathbf{0},j)]^{(0)}+i\epsilon e^\mu_I(\mathbf{0},j)]^{(1)}+\frac{\epsilon^2}{2}[e^\mu_I(\mathbf{0},j)]^{(2)}+...
\label{pereA}
\end{align}
Substituting Eqs. (\ref{perwA}),(\ref{pereA}), and (\ref{perDA}) into Eq. (\ref{eomexpandq}), we have in each order of $\epsilon$:
\begin{align}
\epsilon^0: & \quad 0=\sum_{J\nu}[\mathcal{M}^{\mu\nu}_{IJ}]^{(0)}[e^\nu_J(\mathbf{0},j)]^{(0)} \label{e0}\\
\epsilon^1: &  \quad 0=\sum_{J\nu\xi}[\mathcal{M}^{\mu\nu,\xi}_{IJ}]^{(1)}q^\xi[e^\nu_J(\mathbf{0},j)]^{(0)}+\sum_{J\nu}[\mathcal{M}^{\mu\nu}_{IJ}]^{(0)}[e^\nu_J(\mathbf{0},j)]^{(1)} \label{e1}\\
\epsilon^2: &  \quad  [\omega^{(1)}(\mathbf{0},j)]^2[e^\mu_I(\mathbf{0},j)]^{(0)}=\frac{1}{2}\sum_{J\nu\xi\iota}[\mathcal{M}^{\mu\nu\xi\iota}_{IJ}]^{(2)}q^\xi q^\iota [e^\nu_J(\mathbf{0},j)]^{(0)} \label{e2}\\
&\quad \quad  \quad -\sum_{J\nu\xi}[\mathcal{M}^{\mu\nu,\xi}_{IJ}]^{(1)}q^\xi[e^\nu_J(\mathbf{0},j)]^{(1)}+\frac{1}{2}\sum_{J\nu}[\mathcal{M}^{\mu\nu}_{IJ}]^{(0)}[e^\nu_J(\mathbf{0},j)]^{(2)} . \notag
\end{align}
The LHS in Eqs. \eqref{e0} and \eqref{e1} are set to zero because the acoustic mode vanishes at zero frequency.
For the 0th order, the solution is obvious because of the symmetry properties of  matrix $\mathcal{M}_{IJ}^{\mu\nu}$, listed in the Eq. \eqref{perDA}. We have $[e^\mu_I(\mathbf{0},j)]^{(0)}=\sqrt{m_I}u^\mu(j)$ for arbitrary $u^\mu(j)$. The linear-order equation \eqref{e1} can be written as
\begin{equation}
\sum_{J\nu}[\mathcal{M}^{\mu\nu}_{IJ}]^{(0)}[e^\nu_J(\mathbf{0},j)]^{(1)}=-\sum_{J\nu\xi}[\mathcal{M}^{\mu\nu,\xi}_{IJ}]^{(1)}q^\xi[e^\nu_J(\mathbf{0},j)]^{(0)}. \label{e1b}
\end{equation}
Again using the symmetry properties outlined in \eqref{perDA}, we find that the LHS of Eq. \eqref{e1b} vanishes by multiplying it with $\sum_I\sqrt{m_I}$ on both sides. This condition reduces the number of independent equations for unknown $[e^\nu_J(\mathbf{0},j)]^{(0)}$ to $(N-1)$. In this case, the symmetric matrix $[\mathcal{M}^{\mu\nu}_{IJ}]^{(0)}$ is $(N-1)d\times(N-1)d$, and its inverse $\Gamma^{\mu\nu}_{IJ}$ is also symmetric. Without loss of generality, we can let $[e^\mu_0(\mathbf{0},j)]=0,\mu=1,...,d$ and extend $\Gamma^{\mu\nu}_{IJ}, I,J\neq0$ to $Nd\times Nd$ by letting $\Gamma^{\mu\nu}_{IJ}=0$, if $I$ or $J=0$. Then
\begin{align}
[e^\mu_K(\mathbf{0},j)]^{(1)}&=-\sum_{I=1}^{N-1}\sum_\alpha\Gamma^{\mu\alpha}_{KI}\sum_{J=0}^{N-1}\sum_{\nu\xi}\sqrt{m_J}[\mathcal{M}^{\mu\nu,\xi}_{IJ}]^{(1)}q^\xi u^\nu(j)=-\sum_{J\alpha}\Gamma^{\mu\alpha}_{IJ}\sum_{I\nu\xi}\sqrt{m_I}[\mathcal{M}^{\alpha\nu,\xi}_{JI}]^{(1)}q^\xi u^{\mu}(j).
\end{align}
Given the solutions of 0th and 1st order conditions discussed above, the 2nd order Eq. \eqref{e2} can be written as:
\begin{align}
&\frac{1}{2}\sum_{J\nu}[\mathcal{M}^{\mu\nu}_{IJ}]^{(0)}[e^\nu_{J}(\mathbf{0},j)]^{(2)}=[\omega^{(1)}(\mathbf{0},j)]^2\sqrt{m_I}u^\mu(j)-\frac{1}{2}\sum_{J\nu\xi\iota}\sqrt{m_J}[\mathcal{M}^{\mu\nu,\xi\iota}_{IJ}]^{(2)}q^\xi q^\iota u^\nu(j)\notag\\
&-\sum_{J\alpha\xi}[\mathcal{M}^{\mu\alpha,\xi}_{IJ}]^{(1)}q^\xi \sum_{K\beta}\Gamma^{\alpha\beta}_{IJ}\sum_{L\nu\iota}[\mathcal{M}^{\beta\nu,\iota}_{KL}]^{(1)}\sqrt{m_L}q^\iota u^\nu(j)  \label{e2b}
\end{align}
The calculation leading to Eq. \eqref{e2b}, starting from Eq. (\ref{eomexpandq}), is reproducing the BH method of long waves \citep{Born1952}.
We can still use the properties of the matrix $\mathcal{M}_{IJ}^{\mu\nu}$ as before, so that after multiplying $\sum_I\sqrt{m_I}$ on both sides, the LHS of Eq. \eqref{e2b} vanishes as well. After this operation, we are left with (per volume of unit cell $v_a$):
\begin{align}
&\left(\frac{\sum_Im_I}{v_a}\right)[\omega^{(1)}(\mathbf{0},j)]^2u^\mu(j)=\sum_\nu\left\{\sum_{\xi\iota}[\mu\nu,\xi\iota]q^\xi q^\iota+\sum_{\xi\iota}(\mu\nu,\xi\iota)q^\xi q^\iota\right\}u^\nu(j),
\label{solcond}
\end{align}
where the matrix coefficients in the RHS are described by the notation:
\begin{align}
&[\mu\nu,\xi\iota]=\frac{1}{2v_a}\sum_{IJ}\sqrt{m_Im_J}[\mathcal{M}^{\mu\nu,\xi\iota}_{IJ}]^{(2)}=[\nu\mu,\xi\iota]=[\mu\nu,\iota\xi] \label{m2}\\
&(\mu\xi,\nu\iota)=\frac{-1}{v_a}\sum_{IJ\alpha\beta}\Gamma^{\alpha\beta}_{IJ}\left(\sum_{K}[\mathcal{M}^{\alpha\mu,\xi}_{IK}]^{(1)}\sqrt{m_K}\right)\left(\sum_{L}[\mathcal{M}^{\beta\nu,\iota}_{JL}]^{(1)}\sqrt{m_L}\right)=(\xi\mu,\nu\iota)=(\nu\iota,\xi\mu) . \notag
\end{align}

Comparing Eqs. \eqref{eq9wave2} and (\ref{solcond}), we finally obtain the relation defining the matrix elements of the elastic coefficients:
\begin{align}
&\sum_{\xi\iota}C_{\mu\xi\nu\iota}q^\xi q^\iota=\sum_{\xi\iota}\{[\mu\nu,\xi\iota]+(\mu\xi,\nu\iota)\}q^\xi q^\iota\notag\\
\Rightarrow\quad&C_{\mu\xi\nu\iota}+C_{\mu\iota\nu\xi}=2[\mu\nu,\xi\iota]+(\mu\xi,\nu\iota)+(\mu\iota,\nu\xi)
\label{latticeelastic}
\end{align}
Equation (\ref{latticeelastic}) connects the dynamical matrix and the elastic constants. \textbf{In other words, Eq. \eqref{latticeelastic} obtained in BH framework, provides two paths to elastic constants: lattice vibrations (RHS) or stress-strain relation (LHS). As will be shown in the following sections, the result would be exactly the same as that of the LM formalism. However, as is indicated in Eq. \eqref{e0}, the formula of $C_{\mu\xi\nu\iota}$ is not derived in BH framework. Thus, the way to write down full elastic constant from LHS is not feasible until we report our findings in this paper, which will be shown soon. }

\subsection{The LM formalism} \label{sec2-3}

Nonaffine lattice dynamics has been studied systematically in the Lemaitre-Maloney (LM) formalism, which is applicable to both amorphous materials and ordered crystals. In its framework, the response to external strain is called affine if the interparticle displacements are just the old positions transformed by the macroscopic strain tensor. In a disordered, or a non-centrosymmetric lattice where local inversion symmetry is absent, the situation becomes different since forces from the surrounding environment acting on every particle no longer cancel by symmetry. They have to be relaxed with additional local particle displacements within the unit cell, such that the whole system remains in mechanical equilibrium at every step in the deformation \citep{Milkus2016}. These additional atomic displacements are called nonaffine displacements.

In the language of elasticity, particles (atoms) are assumed to lie in a unit cell described by three Bravais vectors $\underline{\underline{h}}=(\mathbf{a},\mathbf{b},\mathbf{c})$. Thus, the interaction potential depends on both $R_I^\mu$ and $\underline{\underline{h}}$, $\mathcal{U}=\mathcal{U}(R_I^\mu,\underline{\underline{h}})$ and any vector $\mathbf{R}$ is mapped onto a reference cell: $\mathbf{R}=\underline{\underline{h}}\mathbf{w}, w^\nu\in[-0.5, 0.5]$. We call the unit cell before deformation the reference frame $\underline{\underline{\mathring{h}}}$, and denote the deformed cell by the new set $\underline{\underline{h}}$. When a given particle undergoes a displacement to the position $R_I^\mu$, the process can be interpreted in two steps: initially, we have $\mathbf{R}_I=\underline{\underline{F}}\mathring{\mathbf{R}}_I$ where $\underline{\underline{F}}=\underline{\underline{h}}\underline{\underline{\mathring{h}}}^{-1}$ is the deformation gradient tensor. $\underline{\underline{F}}$ describes an affine transformation of the unit cell, whereas $\mathring{\mathbf{R}}_I$ remains unchanged. The external strain $\eta_{\mu\nu}$ defined in Eq. \eqref{eq4strain} is the linear version of the (generalized) Cauchy-Green strain tensor $\underline{\underline{\eta}}=(\underline{\underline{F}}^T\underline{\underline{F}}-\underline{\underline{I}})/2$ that can describes the even nonlinear deformations \citep{Ray1983,Ray1984,Ray1985}. The potential energy can be written either in the reference frame, or in the deformed frame, $\mathring{\mathcal{U}}(\{\mathring{R}_I^\mu\},\underline{\underline{\eta}})\equiv\mathcal{U}(\{R_I^\mu\},\underline{\underline{F}})$. In the second step of the process, particles perform non-affine displacements by relaxing to their nearest equilibrium position $\{R_I^\mu\}$, while the shape of the cell, $\underline{\underline{h}}$ (and hence the tensor $\underline{\underline{F}}$), remains unchanged. Thus, in the reference frame $\{\mathring{R}_I^\mu\}$, changing $\underline{\underline{\eta}}$ means the response to affine strain of the whole system, while the change in the reference configuration $\{\mathring{R}_I^\mu\}$ corresponds to additional nonaffine displacements. Those new coordinates are generally different from the affine positions derived by the reference coordinates, $\{\mathbf{R}_I\}\neq\{\underline{\underline{F}}\mathbf{\mathring{R}}_I\}$. For small deformations, the non-affine equilibrium positions of the particles are a continuous function of $\underline{\underline{h}}:\{R_I^\mu\}=\{R_I^\mu(\underline{\underline{h}})\}$.

When the linear strain is applied slow enough, the deformation can be regarded as quasi-static, and the local mechanical equilibrium is valid at any stage. We can expand the force acting on an individual particle $I$, $f_I^\mu=-\partial\mathring{\mathcal{U}}/\partial\mathring{R}_I^\mu$ (same as in the BH context), in terms of the components of the strain tensor $\underline{\underline{\eta}}$ and $\{\mathbf{R}_I\}$ \citep{Lemaitre2006,Zaccone2011}:
\begin{equation}
0=\delta f_I^\mu=\sum_{J\nu}\frac{\partial^2\mathring{\mathcal{U}}}{\partial\mathring{R}_I^\mu\partial\mathring{R}_J^\nu}\delta\mathring{R}_J^\nu+\frac{\partial^2\mathring{\mathcal{U}}}{\partial R_I^\mu\partial\eta_{\xi\iota}}\delta\eta_{\xi\iota}.
\end{equation}
This is equivalent to the $Nd$ linear system of equations for the nonaffine displacements $\delta\mathring{R}_I^\mu$:
\begin{equation}
\sum_{J\nu}H_{IJ}^{\mu\nu}\delta\mathring{R}_J^\nu=-\Xi_{I,\xi\iota}^\mu\delta\eta_{\xi\iota},  \label{e27}
\end{equation}
where the affine force field $\Xi_{I,\xi\iota}^\mu$ is:
\begin{align}
\Xi_{I,\xi\iota}^\mu=-\frac{\partial^2\mathring{\mathcal{U}}}{\partial\mathring{R}_I^\mu\partial\eta_{\xi\iota}},\quad\text{with \ the \ Hessian}\quad
H_{IJ}^{\mu\nu}=\frac{\partial^2\mathring{\mathcal{U}}}{\partial\mathring{R}_I^\mu\partial\mathring{R}_J^\nu}.  \label{e28}
\end{align}
Clearly, the Hessian matrix defined here is the same as that used in Section 2.2. However, since the LM formalism is applicable to disordered systems, the notation used in \eqref{e28} is slightly different, focusing on the actual particle coordinates rather than displacements from the equilibrium positions in the unit cell. Assuming pairwise interaction, it is easy to see the Hessian matrix is real and symmetric. Hence, it can be diagonalised as $\underline{\underline{H}}=\underline{\underline{P}}\underline{\underline{\Lambda}}\underline{\underline{P}}^T$ where $\underline{\underline{\Lambda}}$ is the diagonal matrix consisting of eigenvalues of $\underline{\underline{H}}$, and $\underline{\underline{P}}$ is the orthogonal matrix whose columns are made of corresponding normalised eigenvectors, $\vec{e}_j,j=1,2,...,Nd$. The $(i,j)$ entry of $\underline{\underline{P}}$ is $e_{ij}$.
Denoting $\delta\hat{\vec{R}}\equiv \underline{\underline{P}}^T\delta\vec{R}$, we have, from transforming Eq.\eqref{e27},
\begin{equation}
\underline{\underline{\Lambda}}\delta\hat{\vec{R}}=-\underline{\underline{P}}^T\vec{\Xi}_{\xi\iota}\delta\eta_{\xi\iota} \label{e29}
\end{equation}
for fixed $\xi\iota$.
Here, vectors originally written in $d$-dimensional space transformed to $Nd$-vectors labeled by an arrow above the symbol. Because of translation invariance, the Hessian matrix contains $d$ zero eigenvalues, so $\underline{\underline{\Lambda}}=diag\{0,...,0,\lambda_{d+1},...,\lambda_{Nd}\}$, where we assume the ordering in eigenvalues without loss of generality. This means, only the components of local displacement $\delta\hat{R}_j, j=d+1,...,Nd$ can be found by solving \eqref{e29}:
\begin{align}\left(
\begin{matrix}
\delta\hat{R}_{d+1}\\
.\\.\\.\\
\delta\hat{R}_{Nd}
\end{matrix}\right)=-\left(
\begin{matrix}
\frac{\vec{e}_{d+1}\cdot\vec{\Xi}_{\xi\iota}}{\lambda_{d+1}}\\
.\\.\\.\\
\frac{\vec{e}_{Nd}\cdot\vec{\Xi}_{\xi\iota}}{\lambda_{Nd}}
\end{matrix}\right).
\end{align}
Transferring back to $\delta R_j$, we obtain
\begin{equation}
\frac{\delta R_j}{\delta\eta_{\xi\iota}}=-\sum_{i=1}^de_{ij}\delta\hat{R}_i-\sum_{i=d+1}^{Nd}e_{ij}\frac{(\vec{e}_i\cdot\vec{\Xi}_{\xi\iota})}{\lambda_i}. \label{e31}
\end{equation}
Here, $\delta\hat{R}_i, i=1,...,d$ are unknown.
The elastic constant is defined as the second derivative of potential energy $\mathring{\mathcal{U}}$ with respect to the strain tensor per unit volume: $C_{\mu\nu\xi\iota}=(\mathcal{D}
^2\mathring{\mathcal{U}}/\mathcal D\eta_{\mu\nu}\mathcal{D}\eta_{\xi\iota})/\mathring{V}$.

Because of mechanical equilibrium, it is easy to verify $\mathcal{D}\mathring{\mathcal{U}}/\mathcal{D}\underline{\underline{\eta}}=\partial\mathring{\mathcal{U}}/\partial\underline{\underline{\eta}}$. Then the elastic modulus is calculated as
\begin{align}
&C_{\mu\nu\xi\iota}=\frac{1}{\mathring{V}}\frac{\mathcal{D}^2\mathring{\mathcal{U}}}{\mathcal{D}\eta_{\mu\nu}\mathcal{D}\eta_{\xi\iota}}=\frac{1}{\mathring{V}}\left(\frac{\partial^2\mathring{\mathcal{U}}}{\partial\eta_{\mu\nu}\partial\eta_{\xi\iota}}+\sum_{I\kappa}\frac{\partial^2\mathring{\mathcal{U}}}{\partial\mathring{R}_I^\kappa\partial\eta_{\mu\nu}}\cdot\frac{\mathcal{D}\mathring{R}_I^\kappa}{\mathcal{D}\eta_{\xi\iota}}\right)
=\frac{1}{\mathring{V}}\frac{\partial^2\mathring{\mathcal{U}}}{\partial\eta_{\mu\nu}\partial\eta_{\xi\iota}}+\frac{1}{\mathring{V}}\sum_{I\kappa}\Xi_{I,\mu\nu}^\kappa\frac{\mathcal{D}\mathring{R}_I^\kappa}{\mathcal{D}\eta_{\xi\iota}}\notag\\
&\equiv C^A_{\mu\nu\xi\iota}+C^{NA}_{\mu\nu\xi\iota}.
\label{e32}
\end{align}
\textbf{The affine moduli, $C^A_{\mu\nu\xi\iota}$, is also called high-frequency moduli that an external constraint (or perturbation) is rapidly transmitted affinely to the system before the relaxational response of the system has a chance to relax fully. The isothermal elasticity tensor is derived in \cite{Tadmor2011}, which reduces to the same form as given in \cite{Lemaitre2006} at $T=0K$. In experiments, the calculated affine moduli is relevant in the high-frequency limit of standard rheological measurements, e.g. shear modulus of glassy polymers in high-frequency oscillatory context \citep{Wittmer2015, Zaccone2013}. Ref. \cite{Wallace1970} compares the bewildering different forms the affine moduli may take if different strain definitions (transformations) are used. This matters especially if external stresses are present. This is the case in virtually all soft matter systems and glasses and also in all systems with internal and external surfaces (surface tension). These stresses contribute linearly to the experimentally relevant small-strain elasticity tensor \citep{Birch1938} and drop out if the energy or free energy is differentiated insisting on a Lagrangian or Eulerian strain. Using the affine terms presented here, a shear modulus of a liquid at a finite pressure would not vanish. This can also be verified using the standard stress-fluctuation formalism \citep{Wittmer2013}. At finite pressured, the presented affine terms are not consistent with the well-known compressed modulus of a standard liquid, as shown via Rowlinson relation \citep{Allen}.}

Using Eq. \eqref{e31}, we can explicitly write the matrix of nonaffine contribution to elastic constants as
\begin{align}
&C^{NA}_{\mu\nu\xi\iota}=\frac{1}{\mathring{V}}\sum_{j=1}^{Nd}\Xi_{j,\mu\nu}\frac{\delta R_j}{\delta\eta_{\xi\iota}}=\frac{1}{\mathring{V}}\sum_{j=1}^{Nd}\Xi_{j,\mu\nu}\left(-\sum_{i=1}^de_{ij}\delta\hat{R}_i-\sum_{i=d+1}^{Nd}e_{ij}\frac{\vec{e}_i\cdot\vec{\Xi}_{\xi\iota}}{\lambda_i}\right)\notag\\
&=-\frac{1}{\mathring{V}}\sum_{i=1}^d(\vec{e}_i\cdot\vec{\Xi}_{\mu\nu})\delta\hat{R}_i-\frac{1}{\mathring{V}}\sum_{i=d+1}^{Nd}\frac{(\vec{e}_i\cdot\vec{\Xi}_{\mu\nu})(\vec{e}_i\cdot\vec{\Xi}_{\xi\iota})}{\lambda_i}   \label{e33}
\end{align}
Because of the structure of $\underline{\underline{\lambda}}$, the (normalized) eigenvectors $\vec{e}_i,i=1,...,d$ all correspond to zero eigenvalues. It suffices to find one case such that they are mutually orthogonal. One simple assignment is to let $e_{ij}=1/\sqrt{N}$ if $j$ is a multiple of $i$, and $e_{ij}=0$ otherwise, for $j=1,...,Nd$. Therefore, the scalar product $\vec{e}_i\cdot\vec{\Xi}_{\mu\nu}=\sum_{j}^{Nd}e_{ij}\Xi_{j,\mu\nu}=1/\sqrt{N}\sum_I^N\Xi_{I,\mu\nu}^i$. In this paper, we only consider pairwise interaction in harmonic approximation, so from Eq. \eqref{e28} the affine force field $\Xi_{I,\mu\nu}^\kappa$ can be expressed as follows:
\begin{align}
&\Xi_{I,\mu\nu}^\kappa=-\sum_{J}\frac{\partial^2\mathring{\mathcal{U}}}{\partial\mathring{R}_{IJ}^\kappa\partial\mathring{R}_{IJ}^\kappa}\frac{\partial\mathring{R}_{IJ}^\kappa}{\partial\eta_{\mu\nu}}
=\sum_{J}\left[(\mathring{R}_{IJ}s_{IJ}-t_{IJ})n_{IJ}^{\kappa}n_{IJ}^\mu n_{IJ}^\nu+\frac{1}{2}t_{IJ}(\delta_{\kappa\mu}n_{IJ}^\nu+\delta_{\kappa\nu}n_{IJ}^\mu)\right]\notag\\
&=\sum_{J}(\mathring{R}_{IJ}s_{IJ}-t_{IJ})n_{IJ}^{\kappa}n_{IJ}^\mu n_{IJ}^\nu,  \label{e34}
\end{align}
with the orientation unit vector $n^\mu$, the tension of a bond $t_{IJ}$ and the stiffness of a bond $s_{IJ}$ defined as
\begin{align}
n_{IJ}^\mu=\frac{\mathring{R}_{IJ}^\mu}{\mathring{R}_{IJ}},\quad t_{IJ}=\frac{\partial\mathring{\mathcal{U}}}{\partial\mathring{R}_{IJ}},\quad s_{IJ}=\frac{\partial^2\mathring{\mathcal
U}}{\partial\mathring{R}_{IJ}^2}.
\end{align}
Here, by $\mathring{R}_{IJ}^\mu$, we mean $\mathring{R}_{IJ}^\mu=\mathring{R}_I^\mu-\mathring{R}_J^\mu$.
To derive the second equality in Eq. \eqref{e34}, we used the identity $\partial\mathring{R}_{IJ}^\kappa/\partial\eta_{\mu\nu}=(\delta_{\kappa\mu}\mathring{R}_{IJ}^\nu+\delta_{\kappa\nu}\mathring{R}_{IJ}^\mu)$. The second term in square brackets vanishes because of the mechanical equilibrium condition, $\sum_Jt_{IJ}n_{IJ}^\mu=0$ for all $I,\mu$.

Due to the inversion symmetry of $\underline{\underline{n}}_{IJ}$ (that is, $\underline{\underline{n}}_{IJ}=-\underline{\underline{n}}_{JI}$), it is clear that $\sum_{IJ}(\mathring{R}_{IJ}s_{IJ}-t_{IJ})n_{IJ}^{\kappa}n_{IJ}^\mu n_{IJ}^\nu=0$, and the first term in Eq. \eqref{e33} vanishes.
Thus, the remaining (negative)  nonaffine contribution  to the elastic constants can be written as
\begin{equation}
C^{NA}_{\mu\nu\xi\iota}=-\frac{1}{\mathring{V}}\sum_{i=d+1}^{Nd}\frac{(\vec{e}_i\cdot\vec{\Xi}_{\mu\nu})(\vec{e}_i\cdot\vec{\Xi}_{\xi\iota})}{\lambda_i}<0 , \label{e36}
\end{equation}
where contributions from zero eigenvalues are excluded in the summation.
Although this derivation and arguments differ from the original LM formalism  \citep{Lemaitre2006,Milkus2016}, the final result of Eq. \eqref{e36} reproduces the key LM result.

\textbf{Accounting for the thermal effect, the Strasbourrg theory shows this affine moduli corrected by nonaffine contribution may be obtained more generally, by averaging different ensembles \citep{Wittmer2013,Wittmer2015}. One can use the integral by parts to reduce the fluctuation of an intensive variable in an ensemble where the average intensive variable is imposed, to a simple average. Imposing a vanishing or finite average intensive varible, one may switch off or or on the Brich coefficients that relates two strains above. Then using the Lebowitz-Percus-Verlet (LPV) transformation between different conjugated ensembles, one can see the complete modulus is given by the affine modulus minus a correction term. As is already shown in \cite{Lutsko1988}, the argument holds in the zero-temperature limit.}

\section{Nonaffine elasticity}\label{sec3}

\textbf{In general, it is cumbersome to apply the BH method directly. One has to express potential energy in terms of Helmholtz displacements, which consist of affine and nonaffine displacements. The affine displacements are related to the external strain, whereas the nonaffine displacements must be solved via Eq. \eqref{Bornequilibrium}, given that one can express the total potential in terms of Helmholtz displacement.} However, we note that, in the BH method, objects like
\[\{\mu\nu\xi\iota\};\left\{\begin{matrix}
I &J\\
\mu &\nu
\end{matrix}\right\};\left\{\begin{matrix}
I &\nu\xi\\
\mu
\end{matrix}\right\},
\]
which appeared, e.g., in the energy density  Eq. \eqref{eq5U2}, are mathematically equivalent to the affine elastic constants $C^A_{\mu\nu\xi\iota}$, the Hessian matrix $H_{IJ}^{\mu\nu}$, and the affine force field $\Xi_{I,\nu\xi}^\mu$, respectively. Therefore,  when we take a derivative of the BH equilibrium condition \eqref{Bornequilibrium} with respect to the strain $\eta_{\mu\nu}$, we do recover the LM condition \eqref{e27}. Since the Hessian always has $d$ zero eigenvalues, and is therefore non-invertible, rather than taking normal mode decomposing and simply ignore zero eigenmodes, one could instead introduce a reduced matrix $\tilde{H}_{IJ}^{\mu\nu}$, and the corresponding local force $\tilde{\Xi}_{I,\xi\iota}^\mu$, by deleting the first $d$ rows and columns in $H_{IJ}^{\mu\nu}$, and the first $d$ elements in $\Xi_{I,\xi\iota}^\mu$, respectively. The reduced $\tilde{H}_{IJ}^{\mu\nu}$ is symmetric and now invertible (see details in the \ref{appC}). With these notations,  the energy density becomes
\begin{equation}
U=\frac{1}{2}\sum_{IJ\mu\nu}\tilde{H}_{IJ}^{\mu\nu}s_I^\mu s_J^\nu+\sum_{I\mu\nu\xi}\tilde{\Xi}_{I,\nu\xi}^\mu s_I^\mu\eta_{\nu\xi}+\frac{1}{2}C_{\mu\nu\xi\iota}^A\eta_{\mu\nu}\eta_{\xi\iota}
\label{eneryreduced}
\end{equation}
which takes the minimum when the local nonaffine displacements $s_I^\mu$ equilibrate:
\begin{equation}
0=\sum_{J\nu}\tilde{H}_{IJ}^{\mu\nu}s_J^\nu+\sum_{\xi\iota}\tilde{\Xi}_{I,\xi\iota}^\mu\eta_{\xi\iota}.
\end{equation}
Solving the minimisation condition for the displacements $ s_I^\mu$, and substituting them back to Eq. (\ref{eneryreduced}), gives the nonaffine correction to the original BH elastic constants:
\begin{align}
&C_{\mu\nu\xi\iota}=C^A_{\mu\nu\xi\iota}-C^{NA}_{\mu\nu\xi\iota}\equiv\frac{1}{V}\frac{\partial^2\mathcal{U}}{\partial\eta_{\mu\nu}\partial\eta_{\xi\iota}}
-\frac{1}{V}\sum_{IJ\kappa\chi}\tilde{\Xi}_{I,\mu\nu}^\kappa(\tilde{H}_{IJ}^{\kappa\chi})^{-1}\tilde{\Xi}_{J,\xi\iota}^{\chi}    \label{e39}
\end{align}
Comparing $C^{NA}$ in Eq. \eqref{e39} with the nonaffine correction in the LM method, Eq. \eqref{e36}, we observe  that these two objects will produce the same result, although they involve different mathematical expressions. Perhaps this is not surprising in retrospect. Here we show the way to obtain these elastic constants via the mathematically well-behaved (invertible)  reduced Hessian matrix and the reduced affine force field, leading to Eq. \eqref{e39}, which we shall call the ``method of reduced fields''.

\begin{figure}
\centering
\includegraphics[width=0.8\textwidth]{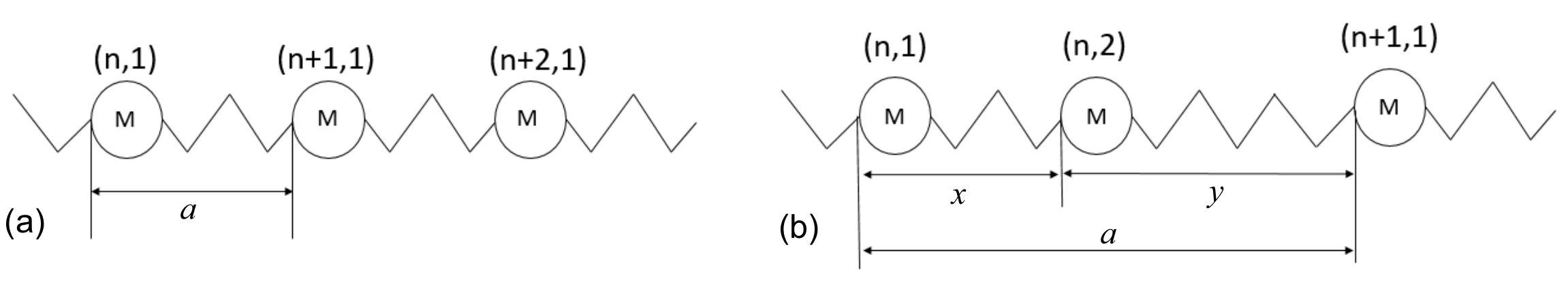}
\caption{\label{fig:chain} Sketch for the lattice examples studied here: (a) 1D linear chain with one mass in each unit cell; (b) 1D linear chain with two masses in a unit cell of size $a$. }
\end{figure}

We will now test all the methods discussed in the previous sections for several specific mechanical models.
First of all, consider the simplest elastic system: the 1D linear chains of equal masses $M$ connected by springs of stiffness $k$, as shown in Fig. \ref{fig:chain}(a). In this case, the potential energy of a deformed string is $\mathcal{U}=\sum_{n}\frac{1}{2}k [R(n+1)-R(n)]^2(1+\eta)^2$, with $\eta$ the imposed strain, and the Hessian matrix being simply a number: $H=2k$. If we want to preserve the lattice periodicity in the disordered state, then there cannot be any nonaffine displacements. The elastic modulus is the same in all three methods: $C=ak$.

\subsection{1D linear chain with two masses in a cell} \label{sec3-1}

Let us now consider a 1D linear chain with two masses connected via springs, all with the same spring constant $k$ but different original lengths, see Fig. \ref{fig:chain}(b). The size of each cell is $a$, within which the original length of spring between masses $m_1$ and $m_2$ is $x$, while $y=a-x$ is the original length of the spring across the neighboring cells. We firstly refer to the BH method of Section \ref{sec2-1}.
After the deformation, positions of masses move from $R_I(n)$ to $R_I(n)+\eta R_I(n)+s_I$, with $I=1,2$ labelling the particle within a unit cell, and $s_I$ the additional nonaffine displacements of each mass in response to the imposed elastic deformation.
The potential energy then takes the form of a sum over cells: $\mathcal{U}=\sum_nV_n$, with
\begin{align}
&V_n=\frac{1}{2}k \left[(1+\eta)(R_2(n)-R_1(n))+s_2-s_1-x\right]^2+\frac{1}{2}k \left[(1+\eta)(R_1(n+1)-R_2(n))+s_1-s_2-y\right]^2. \label{e40}
\end{align}
Note that the spring potentials in $V_n$ reflect the external strain $\eta$ applied.
The internal displacements $s_{1,2}$ are such that the change of potential energy becomes minimal.
Taking $s_1=0$, this minimisation gives
\begin{equation}
s_2=\frac{1}{2}\left[(1+\eta)(R_1(n)+R_1(n+1)-2R_2(n))+x-y\right]
\end{equation}
Substituting $s_{1,2}$ back to $V_n$ and extracting
the quadratic term in $\eta$, we obtain
\begin{equation}
\frac{k}{8}\left[(R_1(n+1)-R_1(n))^2+(R_1(n+1)-R_1(n))^2\right]\eta^2
\end{equation}
Since $a$ is the size of equilibrium repeated cell, and $R_1(n+1)-R_1(n)=a$, the elastic constant from this method is equal to $C=ak/2$.

To check the approach to elastic constants via the collective-modes in Section \ref{sec2-2}, we write the total potential energy within a cell as $V_n= \frac{1}{2}k  [(s_1(n)-s_2(n))^2+(s_2(n)-s_1(n+1))^2]$. Note the expression of $V_n$ is different from Eq. \eqref{e40} because now we focus on lattice vibrations, and do not impose the external strain, and also assume particles are at equilibrium positions initially.
The equation of motion for each mass takes the form
\begin{align}
m_1\ddot{s}_1(n)&=-2k\left[s_1(n)-\frac{s_2(n)+s_2(n-1)}{2}\right]\notag\\
m_2\ddot{s}_2(n)&=-2k\left[s_2(n)-\frac{s_1(n)+s_1(n-1)}{2}\right].
\end{align}
To make it convenient for calculation, we let $m_1=m_2$. The elastic constant can be calculated and gives the same form as in BH method (see details in the \ref{appD}), $C=ak/2$.

To test the LM formalism in Section \ref{sec2-3}, we write potential energy as $\mathcal{U}=\sum_nV_n=k\sum_n[(R_2(n)-R_1(n)-x)^2+(R_1(n+1)-R_2(n)-y)^2]/2$. The energy after putting the (affine) strain $\eta$,
\begin{align}
V_n(\eta)&=\frac{k}{2}[(1+\eta)(R_2(n)-R_1(n))-x]^2+\frac{k}{2}[(1+\eta)(R_1(n+1)-R_2(n))-y]^2
\end{align}
and
\begin{align}
&C^{A}=\frac{1}{a}\frac{\partial^2V_n(\eta)}{\partial\eta^2}=\frac{k}{a}(x^2+y^2)\notag\\
&F_{R_1(n)}(\eta)=-\frac{\partial V_n(\eta)}{\partial R_1(n)}=k[(1+\eta)(R_2(n)-R_1(n))-x]-k[(1+\eta)(R_1(n)-R_2(n-1))-y]\notag\\
&\Xi_{R_1(n)}=\frac{\partial}{\partial\eta}F_{R_1(n)}(\eta)=k(x-y)
\end{align}
where $\Xi_{R_1(n)}$ reflects the affine force acting on each particle.
And similarly,
\begin{align}
&F_{R_2(n)}=k[(1+\eta)(R_2(n)-R_1(n)-x)]+k[(1+\eta)(R_1(n+1)-R_2(n)-y)]\notag\\
&\Xi_{R_2(n)}=k(y-x)
\end{align}
Initial equilibrium condition requires $F_{R_{n,1}}(0)$ and $F_{R_{n,2}}(0)$ are zero, so $R_2(n)-R_1(n)=x, R_1(n)-R_2(n-1)=y$.
The Hessian matrix is
\begin{align}
H=k\left(
\begin{matrix}
&2 &-2\\
&-2 &2
\end{matrix}
\right),
\end{align}
whose eigenvalues are $\lambda_1=0, \lambda_2=4k$, with the eigenvectors corresponding to $\underline{e}_1=1/\sqrt{2}(1, 1)$, $\underline{e}_2=1/\sqrt{2}(1, -1)$. From Eq. (32), we have
\begin{align}
C&=C^{A}-\frac{1}{a}\frac{(\underline{\Xi}\cdot\underline{e}_2)^2}{\lambda_2}=\frac{ak}{2},
\end{align}
where the nonaffine correction is reflected via the $\bm{\Xi}$.
This is consistent with the BH results.\\

Last, we check the different way to find $C^{NA}$, namely if the reduced Hessian and affine force field, as discussed in Eq. (39), can reproduce the correct elastic constant. Deleting the first row and column in $H$ of Eq. (47) and the first element in $\mathbf{\Xi}$, we obtain $\tilde{H}=2k, \tilde{\Xi}=k(y-x)$. The nonaffine correction of Eq. (39) now reads
\begin{equation}
\frac{1}{a}\tilde{\Xi}\tilde{H}^{-1}\tilde{\Xi}=\frac{k(y-x)^2}{2a }.
\end{equation}
With $C^A=(k/a)(x^2+y^2)$, this again gives the correct the elastic constant: $C=ak/2$, but in a faster and more convenient way compared to original BH method.

\subsection{Nonaffinity in non-centrosymmetric lattices} \label{sec3-2}
\begin{figure}
\centering
\includegraphics[width=0.3\textwidth]{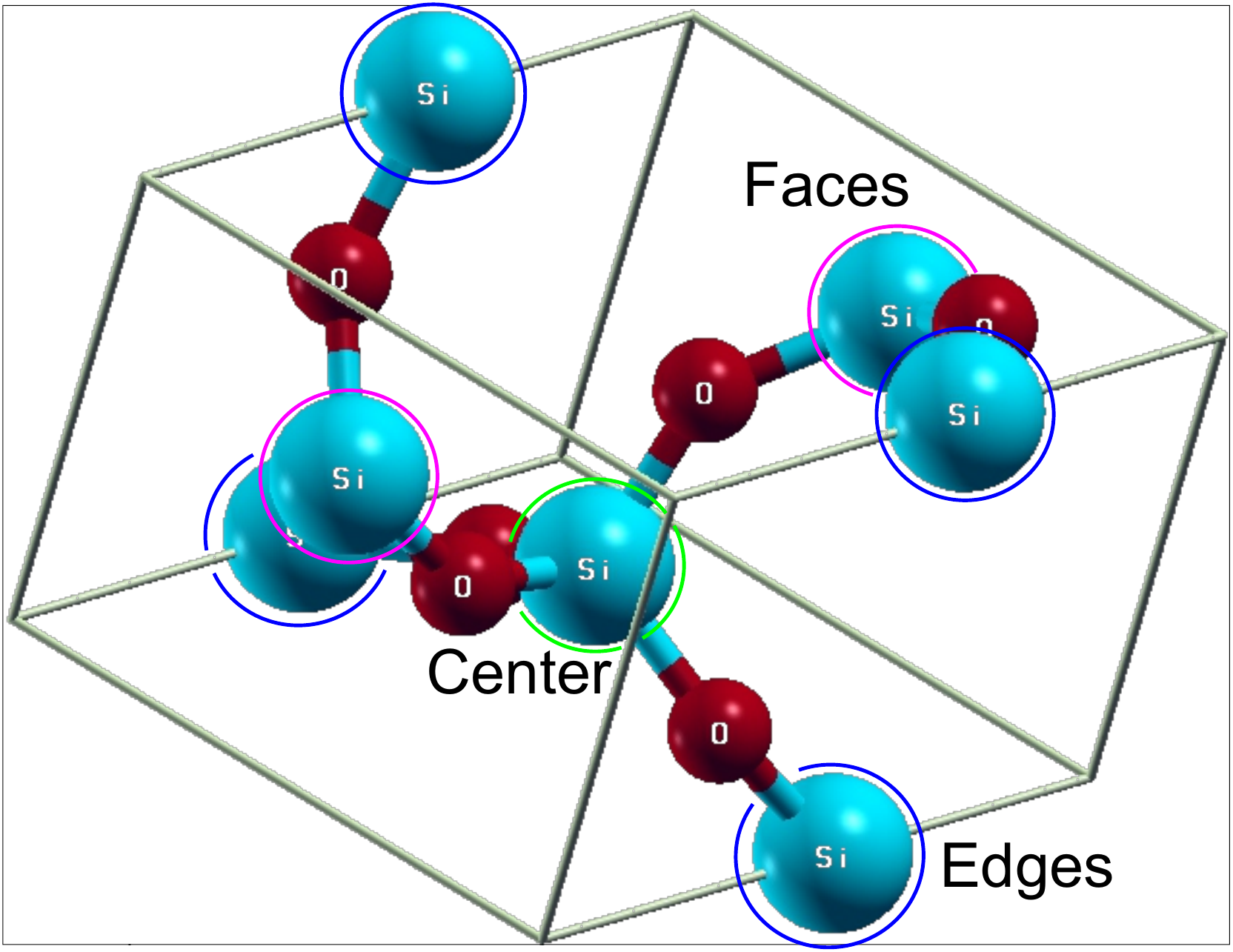}
\caption{Sketch for the unit cell of $\alpha$-quartz, made of the bonded SiO${}_2$ molecules. Si atoms are highlighted: one (green) in the center of the cell; two (purple) on the opposite faces, off center,  and four (blue) on the cell edges. The non-centrosymmetric of such an arrangement gives rise to piezoelectric properties of quartz, as well as to its non-affine deformations. }
\label{quartzview}
\end{figure}

To gain a deeper insight into the original LM formalism and the reduced field method proposed in the paper, we choose a typical non-centrosymmetric lattice system, $\alpha$-quartz, as studied in~\cite{Cui2019nonaffine}. The conventional unit cell, as shown in Fig. \ref{quartzview}, contains three molecules of $\text{SiO}_2$.

The empirical potential is composed of a short-range Buckingham potential plus long-range Coulombic interactions between Si and O atoms \citep{BKS,Mantisi,Carre}. In particular, the short-range potential between atoms $I$ and $J$ is:
\begin{align}
\Phi_{IJ}^{sh}(R_{IJ})=&\left\{A_{IJ}e^{-\frac{R_{IJ}}{\rho_{IJ}}}-\frac{C_{IJ}}{R_{IJ}^6}-\left[A_{IJ}e^{-\frac{R_{c,sh}}{\rho_{IJ}}}-\frac{C_{IJ}}{R^6_{c,sh}}\right]\right\}\times\Theta(R_{c,sh}-R_{IJ}),
\label{BKS}
\end{align}
where $\Theta(R)$ is the Heaviside step function. The cut-off distance is set to be $R_{c,sh}=10\text{\AA}$ to obtain the best agreement with experimental data~\citep{Carre}.
For the Coulombic part, the classical Ewald method was used~\citep{Ewald,Born1952,Toukmaji96,Lee}, with the total electrostatic energy then made of three contributions: short-range term in real space, a long-range term in Fourier space and a self-interaction constant:
\begin{align}
&E\equiv E_{SR}+E_{LR}+E_{SI}\notag\\
&=\frac{1}{4\pi\epsilon_0}\frac{1}{2}\sum
_{I\neq J}\frac{q_Iq_J}{R_{IJ}}\text{erfc}\left(\frac{R_{IJ}}{\sqrt{2}\sigma}\right)+\frac{1}{2V\epsilon_0}\sum_{\mathbf{G}\neq\mathbf{0}}\frac{\exp(-\sigma^2G^2/2)}{G^2}|S(\mathbf{G})|^2-\frac{1}{4\pi\epsilon_0}\frac{1}{\sqrt{2\pi}\sigma}\sum_Iq_I^2,
\label{Etot}
\end{align}
where $q_I$ is the charge on atom $I$, $\text{erfc(z)}$ is the complementary error function, reciprocal lattice vectors are represented by $\mathbf{G}=2\pi[n_x/L_x,n_y/L_y,n_z/L_z]$, and $S(\mathbf{G})=\sum_Jq_J\exp(i\mathbf{G}\cdot\mathbf{R}_J)$ is the structure factor. The dimensions of the simulation cell are labeled as $L_x,L_y,L_z$, which are assumed periodic and orthogonal. The parameter $\sigma$ is the standard deviation of the Gaussian distribution, which is a part of the Ewald method. In the literature, one may also find the use of parameter $\alpha=1/\sqrt{2}\sigma$. The cut-off radius for the real space potential is $R_{cut}=3.12/\alpha=10\text{\AA}$ and a summation in reciprocal space goes up to $n_{\mu,max}=\alpha L_\mu$.
Note that, among these potential energies, the Si-Si longer-range interaction is ignored, because the remaining parts already provide the best agreement with experimental measurements of elastic constants of $\alpha$-quartz~\citep{Carre}. The simulation in~\cite{Cui2019nonaffine} uses a finite system with 1350 atoms in a periodic orthogonal cell. The structure is relaxed at 0K by energy minimisation, followed by adapting the cell dimensions with a barostat to impose zero internal stress, with equilibrium lattice constants obtained.
Knowing the relaxed structure, one can use Eq. \eqref{e32} to compute static elastic constants, some of which are shown in Table I.

\begin{table}[h]
\centering
\begin{tabular}{rcccc}
\hline
Elastic Constants (GPa): &$C_{xx}$ &$C_{yy}$ &$C_{zz}$ &$C_{xy}$ \\ \hline
Affine+Nonaffine:  &90.5 &90.6 &107.0 &41.6
\end{tabular}
\label{tab:elastconst}
\caption{Some values of elastic constants obtained from the LM formalism, and from the reduced-fields method, using the numerical simulation of~\citep{Cui2019nonaffine}. Both approaches produce the same values of constants, within numerical error.}
\end{table}

Alternatively, if one implements the method of reduced fields in Section \ref{sec3}, namely, construct the reduced Hessian $\tilde{H}_{IJ}^{\mu\nu}$ and the affine forced field $\tilde{\Xi}_{I,\xi\iota}^{\mu}$, and compute nonaffine elasticity using Eq. \eqref{e39}, then the same computational procedure as in~\cite{Cui2019nonaffine} can be implemented.  We have thus checked that the same values of elastic moduli are obtained.

\section{Conclusion}
In conclusion, having reviewed different approaches to calculating linear elastic constants, we find that in the BH framework, nonaffine elasticity is essentially due to equilibration of the local additional (Helmholtz) displacements particles experience within the unit cell. In contrast, in the LM formalism, the field of local (Gibbs) forces arising from the breaking of inversion symmetry are instead the cause of nonaffinity. The two methods are equivalent, in the sense that the change of potential energy under the linear strain is minimised, or the mechanical equilibrium condition holds in a conjugate mechanical ensemble. \textbf{In general, the material elasticity tensor is defined through Lagrangian strains, rather than linear strain defined in the paper \citep{Birch1938, Wallace1970}. Experimental small-strain responses differs from affine moduli obtained by differentiation using Lagrangian strains if the system is prestressed. Values obtained using the BH and LM methods only coincide if the system is unstressed.} We also emphasise that, the direct normal mode decomposition to the Hessian matrix always contains zero eigenvalues corresponding to the free translation. Only taking account of non-zero modes in the summation would lead to the correct nonaffine elasticity, which is also what we can get from the suggested method of reduced fields. We thus point out that, the $Nd$ eigenmodes summed to obtain the nonaffine elasticity in \cite{Zaccone2011,Cui2019} should be reduced to $(N-1)d$. The method of reduced field offers an alternative algorithm to study the property of nonaffine vibrational modes. \textbf{The elasticity analysed in this paper is ensemble independent. For systems with different ensembles, the theory can be generalized by averaging over the configurations \citep{Allen}.} Our studies here focus on the periodic lattice, but there is no difficulty for such an analysis to be applied in disordered materials, see \cite{Zaccone2011,Cui2019} for examples.
Only static (equilibrium) elasticity is considered in this paper. It would be of interest to check how these methods are applied for viscoelastic responses, and also compare their validity.

\section*{Acknowledgements}
This work was supported by the CSC-Cambridge Scholarship. Discussions with A. Zaccone are gratefully acknowledged.

\begin{appendix}
\section{Stress-strain relation from homogeneous deformation}\label{appA}
We assume the potential depends on the square of interparticle distance, $V(\mathbf{R}_I(l),\mathbf{R}_J(l'))=V(|\mathbf{R}_I(l)-\mathbf{R}_J(l')|^2)$. The total energy is $\mathcal{U}=\sum_{IJll'}V(\mathbf{R}_I(l),\mathbf{R}_J(l'))/2$. Due to the deformation, the change of square of separation between $\mathbf{R}_J(l')$ and $\mathbf{R}_I(0)$ is
\begin{align}
&|\mathbf{R}_J(l')+\mathbf{s}_J(l')-\mathbf{R}_I(0)-\mathbf{s}_I(0)|^2-|\mathbf{R}_J(l')-\mathbf{R}_I(0)|^2\notag\\
&=\sum_\mu\left(s_J^\mu-s_I^\mu+\sum_\nu s_{\mu\nu}R_{IJ}^\nu(l')\right)^2+2\sum_\mu R_{IJ}^\mu(l')\left(s_J^\mu-s_I^\mu+\sum_\nu s_{\mu\nu}R_{IJ}^\nu(l')\right)
\end{align}
where $s_{\mu\nu}=\partial s^\mu/\partial R^\nu$ is known as deformation parameter.
The corresponding change in $V(\mathbf{R}_I(0),\mathbf{R}_J(l'))$ is the difference of $V$ with the argument of Eq. (A1) and the original $V(\mathbf{R}_I(l),\mathbf{R}_J(l'))$. Then up to the 2nd order, this change is:
\begin{align}
&V'(|\mathbf{R}_{IJ}(l')|^2)\left[\sum_\mu\left(s_J^\mu-s_I^\mu+\sum_\nu s_{\mu\nu}R_{IJ}^\nu(l')\right)^2+2\sum_\mu R_{IJ}^\mu(l')\left(s_J^\mu-s_I^\mu+\sum_\nu s_{\mu\nu}R_{IJ}^\nu(l')\right)\right]\notag\\
&+2V''(|\mathbf{R}_{IJ}(l')|^2)\left[\sum_\mu R_{IJ}^\mu(l')\left(s_J^\mu-s_I^\mu+\sum_\nu s_{\mu\nu}R_{IJ}^\nu(l')\right)\right]^2
\end{align}
Summing these increments of pairwise energy over the unit cell, the change in energy of the unit cell can be written as
\begin{align}
Uv_a&=-2\sum_{I\mu}\left(s^\mu_I+\sum_\nu s^\nu_Is_{\nu\mu}\right)\sum_{l'J}[V'R^\mu]_{\mathbf{R}_{IJ}(l')}+\sum_{\mu\nu}\left(s_{\mu\nu}+\frac{1}{2}\sum_\xi s_{\xi\mu}s_{\xi\nu}\right)\sum_{l'IJ}[V'R^\mu R^\nu]_{\mathbf{R}_{IJ}(l')}+\sum_{IJ\mu\nu}s^\mu_I s^\nu_J\times\notag\\
&\left(\delta_{IJ}\delta_{\mu\nu}\sum_{l'K}[V']_{\mathbf{R}_{IK}(l')}-\delta_{\mu\nu}\sum_{l'}[V']_{\mathbf{R}_{IJ}(l')}
+2\delta_{IJ}\sum_{l'K}[V''R^\mu R^\nu]_{\mathbf{R}_{KI}(l')}-2\sum_{l'}[V''R^\mu R^\nu]_{\mathbf{R}_{IJ}(l')}\right)\notag\\
&-4\sum_{I\mu\nu\xi}s^\mu_Is_{\nu\xi}\sum_{l'J}[V''R^\mu R^\nu R^\xi]_{\mathbf{R}_{IJ}(l')}+\sum_{\mu\nu\xi\iota}s_{\mu\xi}s_{\nu\iota}\sum_{l'JI}[V''R^\mu R^\nu R^\xi R^\iota]_{\mathbf{R}_{IJ}(l')}.
\end{align}
This calculation is reproducing the BH analysis.
Equilibrium conditions require terms linear in $s_I^\mu$ and $s_{\mu\nu}$ vanish, which gives
\begin{align}
&\sum_{l'J}[V'R^\mu]_{\mathbf{R}_{IJ}(l')}=0,\quad \sum_{l'JI}[V'R^\mu R^\nu]_{\mathbf{R}_{IJ}(l')}=0
\end{align}
Eliminating these linear terms, we obtain Eq. (3) in the maintext.

\section{Lattice vibrations and its properties}\label{appB}
Because of periodicity, we have $f^\mu_I(l)=f^\mu_I$ and $H^{\mu\nu}_{IJ}(l l')=H^{\mu\nu}_{IJ}(l-l')$. The whole system has the following properties:\\
(1) Translation invariance: the potential remains unchanged if the system is displaced by an arbitrary vector $\mathbf{e}$, so after expanding the potential to linear order in displacement, we have
\begin{equation}
\sum_{I,\mu}f^\mu_I(l)e^\mu=0 \Rightarrow\sum_If^\mu_I(l)=0,
\label{prop1}
\end{equation}
which is equivalent to $\sum_If^\mu_I=0$.\\
(2) Homogeneous deformation about lattice point $(l, I)$: $s^\mu_I(l)=\sum_\nu s_{\mu\nu}(R^\nu_J(l')-R^\nu_I(l))$. After the deformation, the structure still remains as a perfect lattice. Thus,
\begin{align}
&\sum_I\frac{\partial\mathcal{U}}{\partial s^\mu_I(l)}=0\Rightarrow \sum_I\{f^\mu_I+\sum_{l'J\nu\xi}H^{\mu\nu}_{IJ}(l l')s_{\nu\xi}R^\xi_{IJ}(l-l')+...\}=0\notag\\
&\Rightarrow \sum_{IJl'}H^{\mu\nu}_{IJ}(l-l')R^\xi_{IJ}(l-l')=0.
\label{prop2}
\end{align}
(3) If all particles are displaced from the equilibrium configuration by the same vector $s^\mu_I(l)=e^\mu$, we have
\begin{equation}
\frac{\partial\mathcal{U}}{\partial s^\mu_I(l)}=-f^\mu_I+\sum_{l'J\nu}H^{\mu\nu}_{IJ}(ll')e^\nu+\frac{1}{2}\sum_{l'l''Jk\nu\xi}K^{\mu\nu\xi}_{IJK}(l l' l'')e^\nu e^\xi+...
\label{prop3}
\end{equation}
where coefficients of all orders in $\mathbf{e}$ are zero.\\
(4) If nuclei are displaced by $s^\nu_J(l')=\sum_\xi\omega_{\nu\xi}(R^\xi_J(l')-R^\xi_I(l))=-\sum_{\xi}\omega_{\xi\nu}R^\xi_{JI}(l'-l)$, which is essentially a rotation, then
\begin{equation}
\frac{\partial\mathcal{U}}{\partial s^\mu_I(l)}=-f_I^\mu(l)-\sum_{\xi l'J}H_{IJ}^{\mu\nu}(l-l')\omega_{\nu\xi}R^\xi_{IJ}(l-l')+...
\end{equation}
On the other hand, for a rigid rotation, $\partial\mathcal{U}/\partial s^\mu_I(l)=-\sum_\nu(\delta_{\mu\nu}+\omega_{\mu\nu})f^{\nu}_{I}(l)=-f^{\mu}_I-\sum_{\nu}\omega_{\mu\nu}f^\nu_{I}$. Objects in the bracket after the first equality make the whole term behave as the component of a vector, which corresponds to the transformation matrix for rotation.
Thus, by equating two expressions, we have
\begin{align}
&\sum_\nu\omega_{\mu\nu}f^\nu_I=\sum_{l'J\xi}H_{IJ}^{\mu\nu}(l-l')\omega_{\nu\xi}R^\xi_{IJ}(l-l')=\sum_{lJ\xi}H_{IJ}^{\mu\nu}(l)\omega_{\nu\xi}R^\xi_{IJ}(l)+...
\end{align}
Differentiating both sides with respect to $\omega_{\mu\nu}=-\omega_{\nu\mu}$, we obtain
\begin{equation}
\delta_{\alpha\mu}f^\nu_I+\delta_{\alpha\nu}f^\mu_I=\sum_{lJ}\{H^{\alpha\mu}_{IJ}(l)R^\nu_{IJ}(l)-H^{\alpha\nu}_{IJ}(l)R^\mu_{IJ}(l)\}.
\label{prop4}
\end{equation}

When the nucleus moves to $R_I^\mu(l)+s_I^\mu(l)$, it obeys the equation of motion (e.o.m) $m_I\ddot{s}_I^\mu(l)=-\partial\mathcal{U}/\partial s_I^\mu(l)$. Expanding $\partial\mathcal{U}/\partial s_I^\mu(l)=-f^\mu+\sum_{I'J\nu}H_{IJ}^{\mu\nu}(ll')s_J^\nu(l')$ to the 2nd order, the e.o.m writes
\begin{equation}
m_I\ddot{s}^\mu_I(l)=-\sum_{l'J\nu}H^{\mu\nu}_{IJ}(l l')s^\nu_J(l').
\end{equation}
which is Eq. (10) in the maintext.

The perturbation on $\mathcal{M}^{\mu\nu}_{IJ}(\mathbf{q})$ reads:
\begin{align}
\mathcal{M}^{\mu\nu}_{IJ}(\mathbf{q})&=[\mathcal{M}^{\mu\nu}_{IJ}]^{(0)}+i\sum_{\xi}[\mathcal{M}^{\mu\nu,\xi}_{IJ}]^{(1)}q^{\xi}+\frac{1}{2}\sum_{\xi\iota}[\mathcal{M}^{\mu\nu,\xi\iota}_{IJ}]^{(2)}q^{\xi}q^{\iota}+...\notag\\
[\mathcal{M}^{\mu\nu}_{IJ}]^{(0)}&=\frac{1}{\sqrt{m_Im_J}}\sum_lH^{\mu\nu}_{IJ}(l)=[\mathcal{M}^{\nu\mu}_{JI}]^{(0)}\notag\\
[\mathcal{M}^{\mu\nu,\xi}_{IJ}]^{(1)}&=\frac{-1}{\sqrt{m_Im_J}}\sum_lH^{\mu\nu}_{IJ}(l)R^\xi_{IJ}(l)=-[\mathcal{M}^{\nu\mu,\xi}_{JI}]^{(1)}\notag\\
[\mathcal{M}^{\mu\nu,\xi\iota}_{IJ}]^{(2)}&=\frac{-1}{\sqrt{m_Im_J}}\sum_lH^{\mu\nu}_{IJ}(l)R^\xi_{IJ}(l)R^\iota_{IJ}(l)=[\mathcal{M}^{\nu\mu,\xi\iota}_{JI}]^{(2)}=[\mathcal{M}^{\mu\nu,\iota\xi}_{JI}]^{(2)}
\end{align}
Using Eqs. (\ref{prop1},\ref{prop2},\ref{prop3},\ref{prop4}), one can verify that
\begin{align}
&\sum_{J}\sqrt{m_J}[\mathcal{M}^{\mu\nu}_{IJ}]^{(0)}=\sum_{J}\sqrt{m_J}[\mathcal{M}^{\nu\mu}_{JI}]^{(0)}=0\notag\\
&\sum_{J}\sqrt{m_J}[\mathcal{M}^{\mu\nu,\xi}_{IJ}]^{(1)}=\sum_{J}\sqrt{m_J}[\mathcal{M}^{\nu\xi,\mu}_{JI}]^{(1)}\notag\\
&\sum_{IJ}\sqrt{m_Im_J}[\mathcal{M}^{\mu\nu,\xi}_{IJ}]^{(1)}=0
\label{perDA}
\end{align}

\section{The reduced Hessian and reduced affine force field}\label{appC}
As is explained in section \ref{sec3} in maintext, the energy density $U$ is
\begin{equation}
U=\frac{1}{2}\sum_{IJ\mu\nu}H_{IJ}^{\mu\nu}s_I^\mu s_J^\nu+\sum_{I\mu\nu\xi}\Xi_{I,\nu\xi}^\mu s_I^\mu\eta_{\nu\xi}+\frac{1}{2}C^A_{\mu\nu\xi\iota}\eta_{\mu\nu}\eta_{\xi\iota}
\end{equation}
As is also indicated in maintext, one can arbitrarily assign one $\mathbf{s}_I$ to be zero due to the translation invariance. Without loss of generality, we can set $s_1^\mu=0$. Then the first row and column in $H_{IJ}^{\mu\nu}$ actually contribute nothing in the first term of RHS in Eq. (5). Thus, Eq. (5) can be replaced with Eq. (37) by introducing reduced Hessian matrix and affine force fields. Likewise, looking at the minimisation condition, $\sum_{J\nu}H_{IJ}^{\mu\nu}s_J^\nu+\sum_{\nu\xi}\Xi_{I,\nu\xi}^\mu\eta_{\nu\xi}=0$, this is obviously equivalent to $\sum_{J\nu}\tilde{H}_{IJ}^{\mu\nu}s_J^\nu+\sum_{\nu\xi}\tilde{\Xi}_{I,\nu\xi}^\mu\eta_{\nu\xi}=0$ when $s_1^\mu=0$. For $\tilde{H}_{IJ}^{\mu\nu}$, since the interaction between pairs are uncorrelated in pairwise potential, for each $I$, $\tilde{H}_{IJ}^{\mu\nu}$ are linearly independent for $J,\mu,\nu$. As a result,  $\tilde{H}_{IJ}^{\mu\nu}$ does not have zero eigenvalue and is hence invertible. The minimisation condition requires $s_I^\mu=\sum_{J\nu\xi\iota}(\tilde{H}_{IJ}^{\mu\nu})^{-1}\tilde{\Xi}_{J,\xi\iota}^\mu\eta_{\xi\iota}$. Substituting $s_I^\mu$ back into Eq. (5) leads to Eq. (38) in the maintext.
\section{Application to linear chain with two masses in a cell}\label{appD}
Having equations of motion for each mass:
\begin{align}
m_1\ddot{s}_1(n)&=-2k\left[s_1(n)-\frac{s_2(n)+s_2(n-1)}{2}\right]\notag\\
m_2\ddot{s}_2(n)&=-2k\left[s_2(n)-\frac{s_1(n)+s_1(n-1)}{2}\right]
\end{align}
We try two different ansatz for the solution and check the results separately:\\
1. we assume
\begin{align}
&s_1(n)=\frac{1}{\sqrt{m_1}}e_1(q)e^{i(qna-\omega t)}\notag\\
&s_2(n)=\frac{1}{\sqrt{m_2}}e_2(q)e^{i(qna-\omega t)}
\end{align}
and put them into the coupled equations of motion to get
\begin{align}
\sqrt{m_1}\omega^2e_1=2k\left[\frac{e_1}{\sqrt{m_1}}-\frac{1+e^{-iqa}}{2\sqrt{m_2}}e_2\right]\notag\\
\sqrt{m_2}\omega^2e_2=2k\left[\frac{e_2}{\sqrt{m_2}}-\frac{1+e^{iqa}}{2\sqrt{m_1}}e_1\right],
\end{align}
which is equivalent to
\begin{align}
&(\mathcal{M}(q)-\omega^2I)(e_1\quad e_2)^T=0\notag\\
&\mathcal{M}(q)=\left(
\begin{matrix}
&\frac{2k}{m_1} &-\frac{k}{\sqrt{m_1m_2}}(1+e^{-iqa})\\
&-\frac{k}{\sqrt{m_1m_2}}(1+e^{iqa}) &\frac{2k}{m_2}
\end{matrix}
\right)
\end{align}
For convenience, we let $m_1=m_2$ and from Eq. (23) in the maintext, the coefficient of generic solution $u(j)$ becomes

\begin{equation}
\left(\frac{\sum_{I=1,2}m_I}{a}\right)[\omega^{(1)}(\mathbf{0},j)]^2=\frac{a\kappa}{2}q^2
\end{equation}
and the elastic constant is $C=ak/2$.\\

2. If we instead assume the form of solution to be
\begin{align}
&s_1(n)=\frac{1}{\sqrt{m_1}}e_1(q)e^{i(qna-\omega t)}\notag\\
&s_2(n)=\frac{1}{\sqrt{m_2}}e_2(q)e^{i(qna+qx-\omega t)}
\end{align}
where $x$ is the distance between two masses within one cell. The dynamical matrix becomes
\begin{align}
\mathcal{M}(q)=\left(
\begin{matrix}
&\frac{2k}{m_1} &-\frac{k}{\sqrt{m_1m_2}}(1+e^{-iqa})e^{iqx}\\
&-\frac{k}{\sqrt{m_1m_2}}(1+e^{iqa})e^{-iqx} &\frac{2k}{m_2}
\end{matrix}
\right)
\end{align}
One can define $\tilde{e}_2=e_2e^{iqx}$, then it can be readily to find that the solution is the same as that in 1 and so is the dispersion relation:
\begin{equation}
\omega^2=k(\frac{1}{m_1}+\frac{1}{m_2})\pm k\sqrt{\left(\frac{1}{m_1}+\frac{1}{m_2}\right)^2-\frac{4}{m_1m_2}\sin^2\left(\frac{qa}{2}\right)}
\end{equation}
which is independent of $x$.

Again let $m_1=m_2$, expand $\omega$ up to the 2nd order of wavenumber $k$ and use Eq. (23) in the maintext, the elastic constant is computed to be $C=ak/2$.
\end{appendix}

\bibliographystyle{apalike}
\bibliography{elastic}

\begin{thebibliography}{}

\bibitem[Allen and Tildesley, 2017]{Allen}
Allen, M.~P. and Tildesley, D.~J. (2017).
\newblock {\em Computer Simulation of Liquids: Second Edition}.
\newblock Oxford University Press, Oxford.

\bibitem[Birch, 1938]{Birch1938}
Birch, F. (1938).
\newblock The effect of pressure upon the elastic parameters of isotropic
  solids, according to murnaghan's theory of finite strain.
\newblock {\em Journal of Applied Physics}, 9(4):279--288.

\bibitem[Born and Huang, 1954]{Born1952}
Born, M. and Huang, K. (1954).
\newblock {\em Dynamical Theory of Crystal Lattices}.
\newblock Oxford University Press, Oxford.

\bibitem[Carr{\'{e}} et~al., 2008]{Carre}
Carr{\'{e}}, A., Horbach, J., Ispas, S., and Kob, W. (2008).
\newblock New fitting scheme to obtain effective potential from car-parrinello
  molecular-dynamics simulations: Application to silica.
\newblock {\em Europhys. Lett.}, 82(1):17001.

\bibitem[Cui et~al., 2019a]{Cui2019}
Cui, B., Ruocco, G., and Zaccone, A. (2019a).
\newblock Theory of elastic constants of athermal amorphous solids with
  internal stresses.
\newblock {\em Granul. Matter}, 21(3):69.

\bibitem[Cui et~al., 2019b]{Cui2019nonaffine}
Cui, B., Zaccone, A., and Rodney, D. (2019b).
\newblock Nonaffine lattice dynamics with the ewald method reveals strongly
  nonaffine elasticity of $\alpha$-quartz.
\newblock {\em arXiv preprint arXiv:1908.07982}.

\bibitem[Ewald, 1921]{Ewald}
Ewald, P.~P. (1921).
\newblock The calculation of optical and electrostatic grid potential.
\newblock {\em Ann. Phys. (Leipzig)}, 64:253.

\bibitem[Hoover et~al., 1969]{Hoover1969}
Hoover, W., Holt, A., and Squire, D. (1969).
\newblock Adiabatic elastic constants for argon. theory and monte carlo
  calculations.
\newblock {\em Physica}, 44(3):437 -- 443.

\bibitem[Lacasse et~al., 1996]{Lacasse1996}
Lacasse, M.-D., Grest, G.~S., Levine, D., Mason, T.~G., and Weitz, D.~A.
  (1996).
\newblock Model for the elasticity of compressed emulsions.
\newblock {\em Phys. Rev. Lett.}, 76:3448--3451.

\bibitem[Landau and Lifshitz, 1960]{Landau3}
Landau, L.~D. and Lifshitz, I.~M. (1960).
\newblock {\em Theory of elasticity}.
\newblock Pergamon Press, Oxford.

\bibitem[Langer and Liu, 1997]{Langer1997}
Langer, S.~A. and Liu, A.~J. (1997).
\newblock Effect of random packing on stress relaxation in foam.
\newblock {\em J. Phys. Chem. B}, 101(43):8667--8671.

\bibitem[Lee and Cai, 2009]{Lee}
Lee, H. and Cai, W. (2009).
\newblock Ewald summation for coulomb interactions in a periodic supercell.
\newblock Lecture notes, Stanford University.

\bibitem[Lemaitre and Maloney, 2006]{Lemaitre2006}
Lemaitre, A. and Maloney, C. (2006).
\newblock Sum rules for the quasi-static and visco-elastic response of
  disordered solids at zero temperature.
\newblock {\em J. Stat. Phys.}, 123:415.

\bibitem[Lutsko, 1988]{Lutsko1988}
Lutsko, J.~F. (1988).
\newblock Stress and elastic constants in anisotropic solids: Molecular
  dynamics techniques.
\newblock {\em Journal of Applied Physics}, 64(3):1152--1154.

\bibitem[Mantisi et~al., 2012]{Mantisi}
Mantisi, B., Tanguy, A., Kermouche, G., and Barthel, E. (2012).
\newblock Atomistic response of a model silica glass under shear and pressure.
\newblock {\em Eur. Phys. J. B}, 85:304.

\bibitem[Milkus and Zaccone, 2016]{Milkus2016}
Milkus, R. and Zaccone, A. (2016).
\newblock Local inversion-symmetry breaking controls the boson peak in glasses
  and crystals.
\newblock {\em Phys. Rev. B}, 93:094204.

\bibitem[Radjai and Roux, 2002]{Radjai2002}
Radjai, F. and Roux, S. (2002).
\newblock Turbulentlike fluctuations in quasistatic flow of granular media.
\newblock {\em Phys. Rev. Lett.}, 89:064302.

\bibitem[Ray, 1983]{Ray1983}
Ray, J.~R. (1983).
\newblock Molecular dynamics equations of motion for systems varying in shape
  and size.
\newblock {\em J. Chem. Phys.}, 79(10):5128--5130.

\bibitem[Ray et~al., 1985]{Ray1985}
Ray, J.~R., Moody, M.~C., and Rahman, A. (1985).
\newblock Molecular dynamics calculation of elastic constants for a crystalline
  system in equilibrium.
\newblock {\em Phys. Rev. B}, 32:733--735.

\bibitem[Ray and Rahman, 1984]{Ray1984}
Ray, J.~R. and Rahman, A. (1984).
\newblock Statistical ensembles and molecular dynamics studies of anisotropic
  solids.
\newblock {\em J. Chem. Phys.}, 80(9):4423--4428.

\bibitem[Squire et~al., 1969]{Squirre1968}
Squire, D., Holt, A., and Hoover, W. (1969).
\newblock Isothermal elastic constants for argon. theory and monte carlo
  calculations.
\newblock {\em Physica}, 42(3):388--397.

\bibitem[Tadmor and Miller, 2011]{Tadmor2011}
Tadmor, E. and Miller, R. (2011).
\newblock {\em Modeling materials: Continuum, atomistic and multiscale
  techniques}, volume 9780521856980.
\newblock Cambridge University Press.

\bibitem[Tanguy et~al., 2002]{Tanguy2002}
Tanguy, A., Wittmer, J.~P., Leonforte, F., and Barrat, J.-L. (2002).
\newblock Continuum limit of amorphous elastic bodies: A finite-size study of
  low-frequency harmonic vibrations.
\newblock {\em Phys. Rev. B}, 66:174205.

\bibitem[Toukmaji and Board, 1996]{Toukmaji96}
Toukmaji, A.~Y. and Board, J.~A. (1996).
\newblock Ewald summation techniques in perspective: a survey.
\newblock {\em Comput. Phys. Comm.}, 95:73--92.

\bibitem[van Beest et~al., 1990]{BKS}
van Beest, B. W.~H., Kramer, G.~J., and van Santen, R.~A. (1990).
\newblock Force fields for silicas and aluminophosphates based on ab initio
  calculations.
\newblock {\em Phys. Rev. Lett.}, 64:1955--1958.

\bibitem[Wallace, 1970]{Wallace1970}
Wallace, D.~C. (1970).
\newblock Thermoelastic theory of stressed crystals and higher-order elastic
  constants.
\newblock volume~25 of {\em Solid State Physics}, pages 301 -- 404. Academic
  Press.

\bibitem[Wittmer et~al., 2015]{Wittmer2015}
Wittmer, J., Xu, H., Benzerara, O., and Baschnagel, J. (2015).
\newblock Fluctuation-dissipation relation between shear stress relaxation
  modulus and shear stress autocorrelation function revisited.
\newblock {\em Molecular Physics}, 113(17-18):2881--2893.

\bibitem[Wittmer et~al., 2002]{Wittmer2002}
Wittmer, J.~P., Tanguy, A., Barrat, J.-L., and Lewis, L. (2002).
\newblock Vibrations of amorphous, nanometric structures: When does continuum
  theory apply?
\newblock {\em Europhys. Lett.}, 57(3):423--429.

\bibitem[Wittmer et~al., 2013]{Wittmer2013}
Wittmer, J.~P., Xu, H., Poli{\'{n}}ska, P., Gillig, C., Helfferich, J.,
  Weysser, F., and Baschnagel, J. (2013).
\newblock Compressibility and pressure correlations in isotropic solids and
  fluids.
\newblock {\em The European Physical Journal E}, 36(11):131.

\bibitem[Zaccone and Scossa-Romano, 2011]{Zaccone2011}
Zaccone, A. and Scossa-Romano, E. (2011).
\newblock Approximate analytical description of the nonaffine response of
  amorphous solids.
\newblock {\em Phys. Rev. B}, 83:184205.

\bibitem[Zaccone and Terentjev, 2013]{Zaccone2013}
Zaccone, A. and Terentjev, E.~M. (2013).
\newblock Disorder-assisted melting and the glass transition in amorphous
  solids.
\newblock {\em Phys. Rev. Lett.}, 110:178002.

\end{thebibliography}

\end{document}